\documentclass[sn-mathphys]{sn-jnl}

\jyear{2023}%

\theoremstyle{thmstyleone}%
\usepackage[version=4]{mhchem} 
\theoremstyle{thmstyletwo}%

\theoremstyle{thmstylethree}%

\unnumbered

\begin{document}

\title[High-Temperature Quantum Flux Parametron]{High-Temperature Superconductor Quantum Flux Parametron for Energy Efficient Logic }


\author[1]{\fnm{Han} \sur{Cai}}
\author[2]{\fnm{Jay~C.} \sur{LeFebvre}}
\author[1]{\fnm{Hao} \sur{Li}}
\author[1]{\fnm{Ethan~Y.} \sur{Cho}}
\author[3]{\fnm{Nobuyuki} \sur{Yoshikawa}}
\author*[1]{\fnm{Shane~A.} \sur{Cybart}}\email{cybart@ucr.edu}
\affil[1]{\orgdiv{Electrical and Computer Engineering}, \orgname{University of California, Riverside}, \orgaddress{\city{Riverside}, \postcode{92521 }, \state{CA}, \country{USA}}}
\affil[2]{\orgdiv{Department of Physics}, \orgname{University of California, Riverside}, \orgaddress{\city{Riverside}, \postcode{92521 }, \state{CA}, \country{USA}}}
\affil[3]{\orgdiv{Electrical Engineering}, \orgname{Yokohama National University}, \orgaddress{\city{Yokohama}, \postcode{79-5 Tokiwadai}, \state{Hodogaya}, \country{Japan}}}
\abstract{As we rapidly advance through the information age, the power consumed by computers, data centers, and networks grows exponentially. This has inspired a race to develop alternative low-power computational technologies. A new adiabatic configuration of a decades-old superconducting digital logic device has darted into the lead called quantum flux parametrons (QFP). QFP operate with dissipation so low that they seemingly violate the laws of thermodynamics. In just a short span of time, they have gone from simple single NOT gates to complex processors containing thousands of gates. They are fabricated from elemental niobium superconductors cooled to just a few degrees above absolute zero. However, their efficiency is so great that for large high-performance computers with several gates, the energy savings are immense. For smaller computational platforms QFPs from high-temperature superconductors (high-\emph{T}$_\text{C}$) are highly desirable. In this work, we take the first steps towards this goal with the demonstration of a high-\emph{T}$_\text{C}$ QFP shift register. Our device is fabricated using focused helium ion beam lithography where the material is modified with an ion beam at the nanoscale to directly pattern these circuits into a high-\emph{T}$_\text{C}$ thin film. We validate the correct logical operation at 25~K, over 6 times higher than niobium devices with an estimated bit energy of 0.1 attoJoule at 10~GHz.
}



\maketitle

\section{Main}\label{sec1}
The quantum flux parametron (QFP) is a type of superconducting digital logic circuit invented by Goto in the late 1980s\cite{HaradaQFP}. It did not receive much attention at the time as the superconducting electronics (SCE) research community mainly focused on a much faster type of SCE device called rapid single flux quantum (RSFQ) logic, that can operate at clock frequencies of hundreds of GHz\cite{LikharevRSFQ, chen1999rapid}. It was once thought that SFQ would replace semiconductors due to these high computational speeds however innovation in silicon complementary metal oxide semiconductor (CMOS) logic performance, driven predominantly by scaling, soon out-paced SCE. 

Today, we are facing a new challenge in that the associated energy consumption of computer chips is growing exponentially and could eventually outpace our global supply of energy \cite{landauer1961irreversibility,  shizume1995heat,  sagawa2009minimal}. This energy deficit will be exacerbated by the impending transition from fossil fuels to renewable sources. This motivates the need for new low-energy approaches to how chips function and has renewed interest in low-power superconducting digital logic \cite{bennett1973logical, semenov2003negative, ren2011progress, wustmann2020reversible, semenov2023biosfq}. 

\begin{figure}[htbp]
\centerline{\includegraphics[width=0.8\columnwidth]{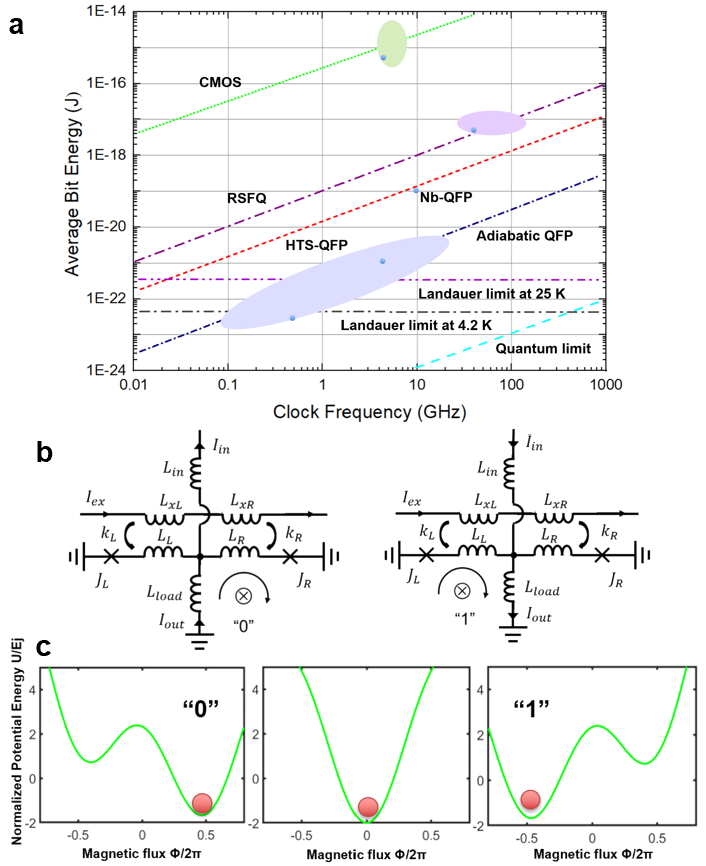}}
\caption{Quantum flux parametron logic  (a) shows a comparison of bit energy for several digital logic circuit families highlighting the low power of QFP \cite{volkmann2012implementation,  vernik2014design, takeuchi2015thermodynamic, takeuchi2022adiabatic}. QFP circuit diagrams (b) and potential energy (c) for the "0" and "1" logic states.
 }
\label{fig:fig1} 
\end{figure}

In 2013 Takeuchi et al published a deep analysis of the energy dissipated in the QFP and found that with a particular choice of circuit parameters \cite{takeuchi2013adiabatic}, QFP could operate at power levels of 12~yJ \cite{takeuchi2013simulation} which is over 6 orders of magnitude lower than CMOS and two orders of magnitude lower than SFQ logic (Fig.\ref{fig:fig1}(a)). This is because the switching between the logic states of "0" and "1" can be done smoothly and efficiently unlike the rapid switching events in other logic. They describe this low energy configuration of QFP as adiabatic or AQFP to liken it to reversible processes in thermodynamics such as the Carnot cycle. This is because AQFP can be configured for reversible computing that operates at energies below the fundamental Landauer limit from information theory \cite{takeuchi2014reversible, chen2019adiabatic}. The Landauer limit at 4.2~K is an energy level of $4.019 \times 10^{-23}$ J that is derived from the fundamental entropy of two logic states ($S=k_bT\text{ln}2$) \cite{landauer1961irreversibility}. However, in reversible computing\cite{fredkin1982conservative,bennett1973logical} entropy does not increase because the logic is symmetric meaning that the answer can be fed back through the circuit to generate the question. This removes the fundamental limit on how low the dissipation energy can go. Before the development of AQFP, the field of reversible computing was predominately a theoretical one with many proposed devices\cite{likharev1977dynamics, semenov2003negative, ren2011progress} and experimental demonstrations limited to processes in biological molecules\cite{klein1999biomolecular} and quantum dots\cite{lent1997device}. Now however it is possible to build and test experimental AQFP gates that operate at the lowest energies of any logic device.

A decade later AQFP is now the reining champion of all low-power digital logic. The technology has rapidly advanced from the simple circuits featured in that first paper\cite{takeuchi2013adiabatic} to processors containing tens of thousands of gates \cite{ayala2020mana}. Design rules and cell libraries have been established for the fabrication of AQFP circuits \cite{takeuchi2017adiabatic, he2020compact} and several research groups around the world have been designing and building AQFP chips \cite{tadros2019systemverilog, lee2021irredundant, luo2022scalable, Takeuchi2023scalable}. 

Fig.\ref{fig:fig1}a shows the comparison of energy dissipation between different digital logic circuits. AQFP is the lowest and dips below the Landauer limit as frequency decreases. The devices that we will demonstrate in this article fall on the line directly above it, labeled HTS-QFP. 
Fig.\ref{fig:fig1}b shows electrical schematics for QFP circuits and the potential energy landscape for the "0" and "1" logic states in Fig.\ref{fig:fig1}c. It shows that if an input signal flows from the ground  through the load inductor and corresponds to an excitation pulse the potential well on the right deepens and a flux quantum is stored in the circuit. If the input signal is reversed the left side potential well becomes lower energy and the flux quantum shifts to the other side or state. Careful selection of the inductor values can be done to make the switching event very low energy and adiabatic at a relatively high speed.

Nearly all SCE are fabricated using niobium Josephson junctions because of their unmatched reproducibility and relative ease of fabrication. However, the low superconducting transition temperature of elemental Nb (9.3~K) requires large power-hungry (10s of kW)\cite{ter2002low}  refrigeration systems or liquid helium. This requirement limits use to very large platforms and is hardly portable. In contrast, high-transition temperature (high-$T_\text{C}$) ceramic superconductors like \ce{YBa2Cu3O7} (YBCO) can operate on much smaller lower power (10s of watts)\cite{ter2002low} coolers that can fit in a rack-mountable enclosure around the size of a server.

Unfortunately, high-$T_\text{C}$ materials are very difficult to process into devices with reproducible and predictable electrical characteristics. A promising approach to realizing YBCO circuits is to use a focused helium ion beam (FHIB)\cite{ward2006helium} to create nanoscale planar YBCO Josephson junctions. In 2015, Cybart et al. \cite{cybart2015nano} demonstrated this process and showed that a 0.5~nm diameter focused helium ion beam could be used to convert the superconducting material to an insulator at the nanometer scale\cite{wang2019estimation} to directly write Josephson tunnel junctions, superconducting nanowires\cite{cho2018superconducting}, SFQ devices\cite{cai2021yba} and nanoscale magnetic sensors\cite{li2020high}. 
In this work, we combine these components to create nanoscale high-$T_\text{C}$ QFP logic and demonstrate a three QFP shift register with an estimated bit energy of $\sim$ $1 \times 10^{-19}J$ at 10~GHz. Using FHIB nanofabrication to create (high-$T_\text{C}$) QFP has great potential in highly scalable low-energy digital devices. High-$T_\text{C}$ superconductors have much higher critical current density than niobium that, allows Josephson junctions to be about 100 times smaller in area while the planar geometry afforded by the FHIB process will enable them to be more densely packed than conventional Nb junctions. Furthermore, the larger sheet inductance of high-$T_\text{C}$ allows the flux storage loops to be about 15 times smaller. Adding up all of these factors could allow for $10^3$ more gates in high-$T_\text{C}$ materials than the current state-of-the-art niobium for the same areas. 

\begin{figure}[htbp]
\centerline{\includegraphics[width=0.9\columnwidth]{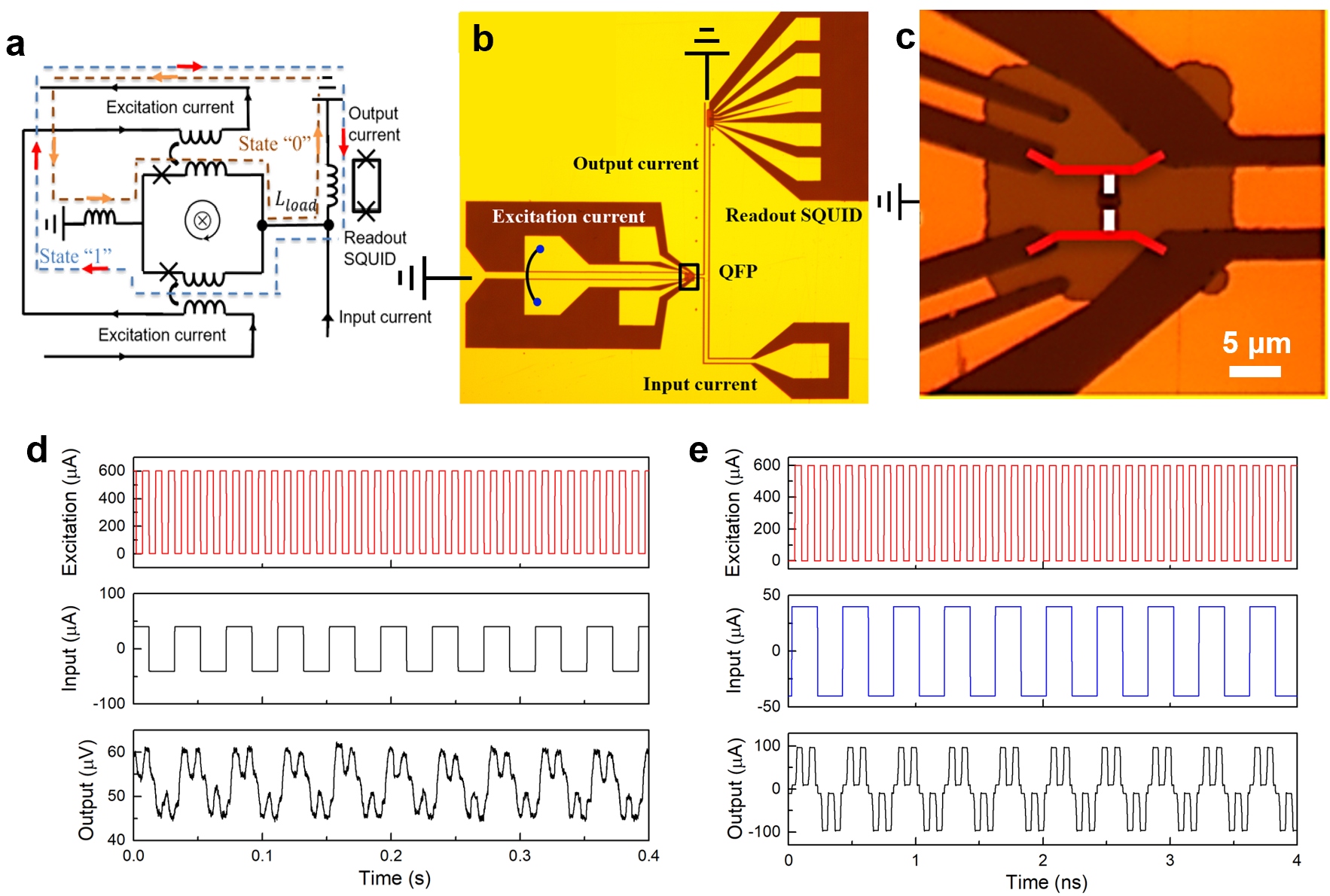}}
\caption{YBCO Quantum Flux Parametron (a) Circuit schematic for the YBCO QFP, highlighting the current paths for storage of flux quanta representing the "0" and "1" logic states. (b) Optical photograph of the QFP input transmission line and readout SQUID. (c) The optical photograph of the exposed YBCO pattern on an expanded scale, where the gold electrode layer was chemically etched away. The solid white lines represent the Josephson junctions directly written into the YBCO layer by the FHIB irradiation. The red lines are insulating lines written to isolate the QFP from the excitation current. (d) The operation of the QFP switching gate was measured at 25~K. A steady output voltage was observed with 600~$\mu$A excitation current and $\pm$40~$\mu$A input current at clock 100~Hz. (e) Simulation of the QFP operating at 10~GHz using the experimental circuit parameters.}
\label{fig:QFP} 
\end{figure}

To create high-$T_\text{C}$ QFP, a circuit was designed that could be implemented in a single layer of the superconductor to avoid the need to fabricate vias and crossovers (Fig.\ref{fig:QFP}a). The starting material was a 40~nm thick YBCO film grown by reactive co-evaporation on a sapphire wafer\cite{wang2019yba}. A 200~nm thick gold electrode was deposited on top inside of the same chamber without breaking the vacuum to create a low-resistance ohmic electrical contact. After dicing the wafers into $5 \times 5$~mm chips, laser lithography was used to pattern the circuit features shown in (Fig.\ref{fig:QFP}b). These included a coplanar transmission line for input, a superconducting quantum interference device (SQUID) for readout, and the QFP cell. Fig.\ref{fig:QFP}c) is a zoomed region of the QFP cell. A subsequent lithography and chemical etching step is used to remove the gold contact and uncover the buried YBCO for FHIB patterning. The two vertical white lines represent where the beam is scanned to create two Josephson junctions. The two red lines are where the FHIB is used to create insulating regions to isolate two excitation coils from the QFP. We estimate these lines to be of the order of $\sim$~10~nm wide which results in strong inductive coupling of the excitation signal\cite{li2020high} to the QFP. 

The QFP was cooled to 25~K and operated with a $\pm$40~$\mu$A input current at 100~Hz. The result is shown in (Fig.\ref{fig:QFP}d). The output of the readout SQUID tracks with the input but switching between the states only occurs after a falling edge from the excitation. A steady output voltage was observed with 600~$\mu$A excitation current. A dc-SQUID biased in the voltage state is used as a current sensor to measure the output current flowing through $L_\text{Load}$. The load inductance of the QFP is magnetically coupled to the dc-SQUID and the dc-SQUID exhibits a voltage pulse when the QFP switches to a $"0"$ or $"1"$ state.  The operation point of the dc-SQUID is set to the maximum transfer factor $\mid$ $\frac{dV}{dI_B}$ $\mid$ point ($I_B$ is the magnetic current in the SQUID loop), where it is the most sensitive to external magnetic current and can continuously and linearly convert the output current of the QFP into an output voltage. Due to the planar nature of the device, we detect a dither on the output from parasitic coupling from the excitation signal. However, it does not interfere with the correct operation. The low test frequency is due to the limitations of our cryostat wiring. Simulations were performed using WRSpice using the experimental properties of the QFP and are shown in (Fig.\ref{fig:QFP}e). They confirm the correct operation and show that the QFP chip can operate as fast as 10~GHz.

\begin{figure}[htbp]
\centerline{\includegraphics[width=0.9\columnwidth]{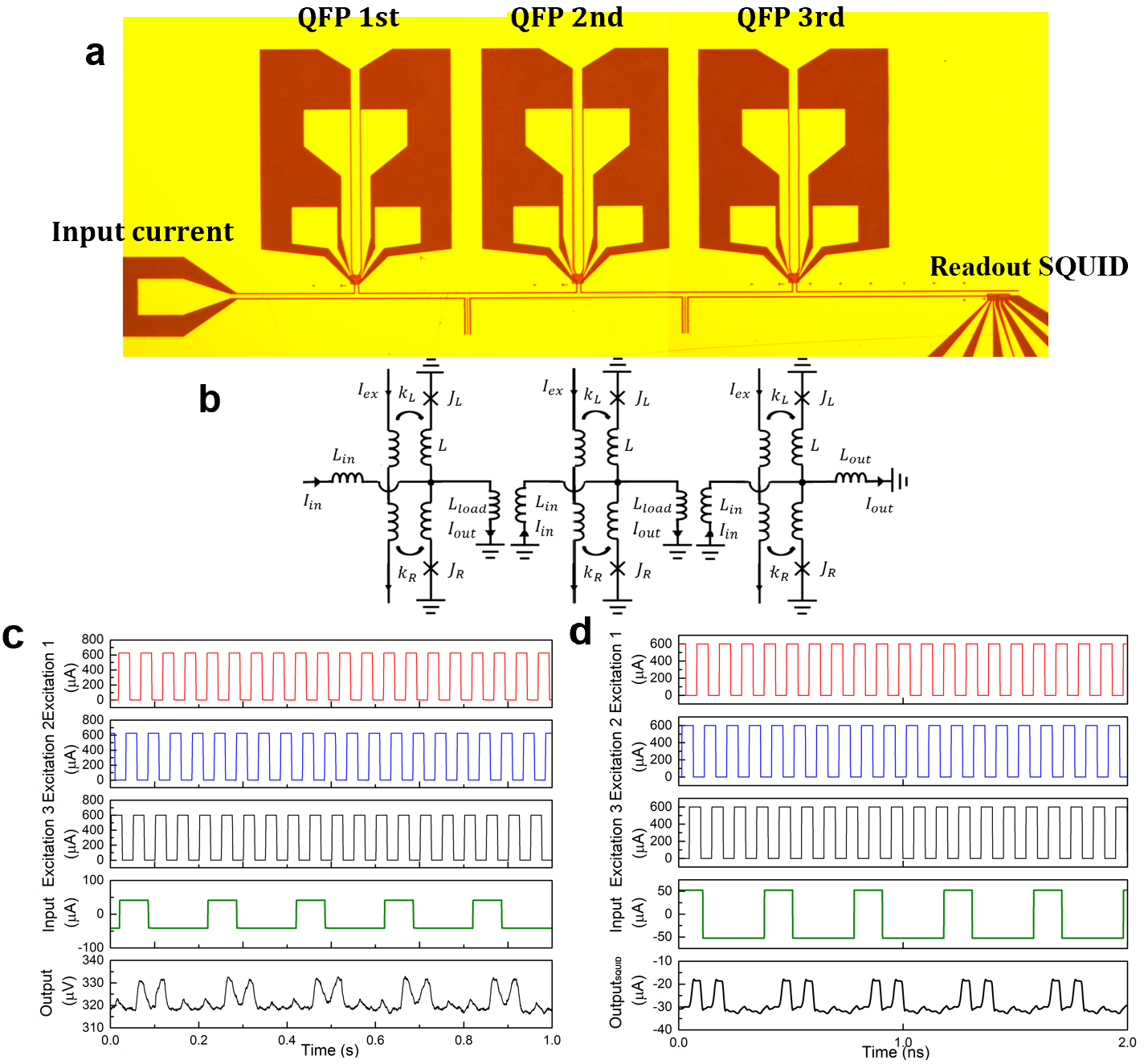}}
\caption{(a) The optical microscopic photograph  of the three-QFPs array. (b) The equivalent logic circuit of the three-QFPs array. (c) The experimental operation of the three-QFPs array.  (d) Transient analysis of the three-QFPs array operating at 10~GHz. }
\label{fig:QFParray} 
\end{figure}

To conduct a more stringent test of correct operation and timing a three-stage QFP shift register was fabricated and shown in Fig.\ref{fig:QFParray}a. In this circuit, three QFP cells were inductively coupled to the input line. The equivalent circuit schematic is shown in Fig.\ref{fig:QFParray}b. The excitation lines of the QFP array were driven by a three-phase clock whose phases are shifted $120^\circ $ relative to each other. As shown in Fig. \ref{fig:QFParray}c, the excitation currents are activated in turn. When the input current is positive, each QFP generates an output current after each excitation current is activated, which propagates one bit of information. An output voltage is obtained by the readout SQUID when the third clock is in the high state. WRSpice was used to simulate this shift register, and the simulation shown in Fig.\ref{fig:QFParray}d agrees well with the measured result. 

These demonstrations are the first step to a new family of QFP logic devices from high-$T_\text{C}$ superconductors operating at a temperature six times higher than prior-art devices. The scalability is unmatched by any other superconductor technology and the power consumption is only bested by low-temperature niobium AQFP. We envision it as a technology that can be integrated with other superconductor technologies as well as CMOS to provide ultra-low energy portable superconducting circuits for high-performance computing.  

\section{Methods}
\subsection{Device Fabrication}
A commercial 40-nm-thick YBCO thin film was grown by Ceraco GMBH by reactive co-evaporation on cerium oxide buffered r-plane sapphire. For electrical contacts on the YBCO chip, a subsequent 200-nm-thick Au layer was thermally evaporated in situ. Laser photolithography was used with Fuji Film OCG825 positive photoresist to define the electrodes. Argon ion milling was performed in an Ion tech 21cm broad beam Kaufman source at 500V.  A second photolithography and KI-I chemical etch were used to remove the Au layer and uncover the YBCO area for FHIB direct writing.  Helium ion irradiation was accomplished by a Zeiss Orion NanoFab He/Ne ion microscope with 32~keV He$^+$ ions. An ion dose of $6 \times 10^{16}$ He$^+/ $cm$^2$ was applied to create a very narrow insulating barrier for YBCO in-plane superconducting-insulating-superconducting (SIS) junctions. A dose of $3 \times 10^{17}$ He$^+/ $cm$^2$ was used to create the insulating lines between the adjacent excitation current line and the QFP loop. 

\subsection{Measurement}
The devices were cooled in an evacuated dip probe with a $\mu$-metal magnetic shield vacuum can in a liquid helium Dewar. The height of the probe in the bath was adjusted to set the temperature to $\sim$ 25~K. An arbitrary function generator generated the input and excitation currents with a 10~k$\Omega$ resistor placed at each input terminal to suppress voltage noise from room temperature to low temperature. Low-noise battery-powered pre-amplifiers were used to measure the output voltage. All the instruments and Dewar were in an RF-shielded room and connected to a general ground to reduce noise. The dip probe was wired with low-frequency twisted pair manganin wires which limited the frequency of operation.  

\subsection{Simulation}
This circuit is simulated by WRspice using  the experimental circuit parameters. The QFP gates are driven by the excitation clock pulses with an amplitude of 600 $\mu$A and an operating frequency of 10~GHz. The inductance values are extracted from the circuit layout by InductEx: k $\approx$ 0.4,  $\beta_{L}\approx1.1$, $\beta_{Load}\approx 3.5$ and are comparable to experimental measurements in the same films\cite{li2019inductance}. Here $\beta_{L}$ and $\beta_{Load}$ are the normalized inductances of L and $L_\text{Load}$, respectively. For the high-speed operations, critically damped Josephson junctions with the McCumber parameter $\beta_c\approx 1$ are used in this circuit.  A small bidirectional input current $I_\text{in}$ is set to be 40 $\mu$A. When the excitation current is activated, the QFP gate switches at every cycle of the input current and generates an output current. The different polarity of the output current represents the different logic states of the QFP gate. Compared to the input current, the output current is amplified with a current gain $I_\text{in}$/$I_\text{out}$ about 2.5, which is a feature of the QFP gate.

\section{Supplemental Material}\label{sec12}
 A small offset was observed in the output voltage of Fig. \ref{fig:QFP}d. Two effects might induce this offset. One is the parasitic coupling in the actual asymmetrical construction of the circuit layout, causing  unwanted magnetic coupling between the excitation inductance and load inductance $L_\text{Load}$, which then is read out by the dc-SQUID. If this parasitic magnetic coupling is not small enough, the excitation current margin is reduced. Direct measurements of the mutual inductance were obtained by directly driving the SQUID by the potential parasitic coupling lines to investigate the influence of the parasitic coupling. The mutual inductance can be obtained by the relation $M=\phi_0/\triangle I_B$. $\triangle I_B$ is the period of magnetic current that produces a single flux quantum in the SQUID loop. As shown in Sup.Fig.\ref{fig:coupling}a, the parasitic inductance is about 0.04~pH. A 600~$\mu$A excitation current will produce about 0.2~$\mu$V offset in the output voltage. This offset resulted in a $\sim$0.2~$\mu$V difference between the output voltage amplitude of state "0" and state "1", which is negligibly small. The input current is the other effect that would induce an offset in the output voltage. As shown in Sup.Fig.\ref{fig:coupling}b,  a 40~$\mu$A (-40~$\mu$A) input current will cause a $\sim$2~$\mu$V (-2~$\mu$V) offset. This offset corresponds to a $\sim$4~$\mu$V offset between two polarity pulses in Fig. \ref{fig:QFP}d. Generally, the input current of the QFP gate can be much smaller than 40~$\mu$A, and the input current's influence will be suppressed. Moreover, $\sim$90~$\mu$A is needed from voltage magnetic current characteristics to produce a 5~$\mu$V modulation corresponding to the output current. Therefore, the experimental current gain of the QFP gate is about 2.25.

\renewcommand{\figurename}{Sup. Fig.}
\setcounter{figure}{0}
\begin{figure}[htbp]
\centerline{\includegraphics[width=1.1\columnwidth]{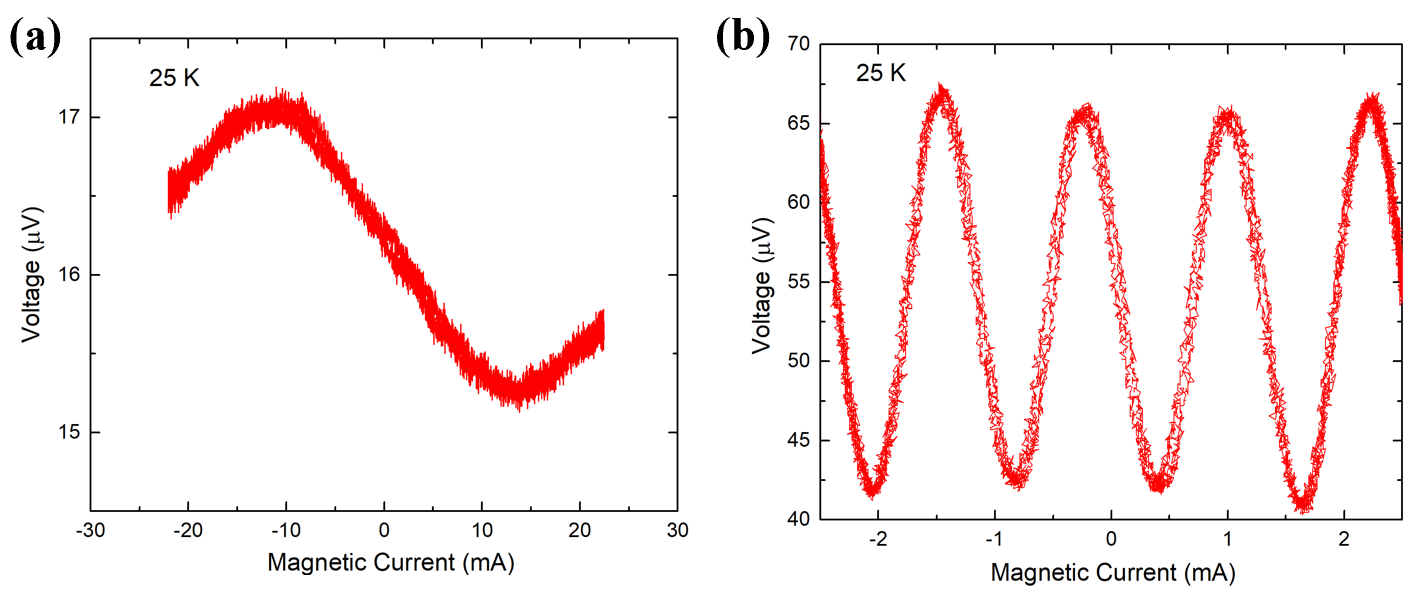}}
\caption{Voltage magnetic current characteristics for (a) readout SQUID modulated by the excitation current and (b) readout SQUID directly modulated by the input current. }
\label{fig:coupling} 
\end{figure}

\backmatter

\bmhead*{Acknowledgments} 
This work was partly supported by the Air Force Office of Scientific Research under Grants FA9550-20-1-0144.


\bibliography{sn-bibliography}




\end{document}